\documentclass[rnote,traditabstract]{template/aa} % for the research notes
\usepackage{template/natbib}
\usepackage{array}
\usepackage{floatrow}
\usepackage{float}
\usepackage[fleqn]{amsmath}
\usepackage{amssymb}
\usepackage{stfloats}
\usepackage{morefloats}
\usepackage{multirow}
\bibpunct{(}{)}{;}{a}{}{,}
\usepackage[dvips]{graphicx}
\usepackage{slashbox}
\usepackage{xfrac}
\usepackage{hyperref}
\usepackage{breakurl}
\usepackage{txfonts}
\usepackage{breqn}

\usepackage{pstricks}
\usepackage{caption}
\usepackage{subcaption}
\usepackage{color}
\usepackage{breqn}
\usepackage[shortcuts]{extdash}
\usepackage{cases}
\usepackage{tikz}
\begin{document}
  \title{Deep {\it Herschel}\thanks{{\it Herschel} is an ESA space observatory 
   with science instruments provided by European-led Principal Investigator consortia and with important 
   participation from NASA.} PACS point spread functions\thanks{FITS files of our PACS PSFs (Fig.~\ref{fig:mosaic}) are 
   available in electronic form at the CDS via anonymous ftp to cdsarc.u-strasbg.fr (130.79.128.5)
or via \url{http://cdsweb.u-strasbg.fr/cgi-bin/qcat?J/A+A/}}}

   \author{M. Bocchio\inst{1}, 
          S. Bianchi\inst{2}
          A. Abergel\inst{1}}
          
   \institute{Institut dÕAstrophysique Spatiale (IAS), UMR8617, CNRS, Universit\'e Paris Saclay, Universit\'e Paris Sud, Orsay F-91405, France\\
        INAF - Osservatorio Astrofisico di Arcetri, Largo Enrico Fermi 5, 50125 Firenze, Italy\\
                                }
 
  \abstract
  {The knowledge of the point spread function (PSF) of imaging instruments represents a fundamental requirement for astronomical 
  observations.
  The {\it Herschel} PACS PSFs delivered by the instrument control centre are obtained from observations of the Vesta asteroid,
  which provides a characterisation of the central part and, therefore, excludes fainter features.
  In many cases, however, information on both the core and wings of the PSFs is needed.
  With this aim, we combine Vesta and Mars dedicated observations and obtain PACS PSFs with an unprecedented dynamic range ($\sim 10^{6}$)
  at slow and fast scan speeds  for the three photometric bands.
}


   \maketitle
%
%________________________________________________________________

\section{Introduction}

%For radio telescopes and diffraction-limited space telescopes the dominant terms in the PSF may be inferred from the configuration of the aperture in the Fourier domain.

The response of a given imaging instrument to a point source is known as the point spread function (PSF).
In the case of diffraction-limited space telescopes this quantity is dominated by the configuration of the aperture
and it is key to many aspects of astrophysical observations.
First, models are often compared to observations. This operation is typically carried out by convolving
a given model to the PSF of the observed image. An error in the estimate of the PSF would lead to errors
in the interpretation of the observations.
Second, images taken with different instruments or at different wavelength bands intrinsically 
have distinct resolutions.
In order to compare multiple images on a pixel-by-pixel basis they need to be smoothed to a common
(larger) PSF, which is a procedure that involves the use of convolution kernels.
The construction of a convolution kernel is based on the knowledge of the PSF of the image that 
needs to be processed and the common PSF.
Third, the interface between different regions (e.g. photodissociation regions or the outskirts of galaxies) 
are often rich in information. A strong gradient in intensity usually characterises 
the interface, however, making the contrast between these regions very strong.
Faint wings of the PSF of the instrument can extend very far from the PSF centre and 
%, e.g. structures at $10^{-4}$ with respect to the peak 
%are present at $\sim 150''$ from the centre of the PACS 160 $\mu$m PSF (FWHM $\sim 12''$).
if the contrast between regions is sufficiently high, faint structures of the PSF can have an intensity that is comparable to that of the fainter regions.
This represents a possible source of contamination and needs to be carefully taken into account.

The PSF of the photometer of the PACS instrument \citep{2010A&A...518L...2P} on board {\it Herschel} is characterised by (\citealt{PICC}) a narrow core, 
a tri-lobe pattern, and knotty structured diffraction `rings' at sub-percent level. 
For fast scans in standard and parallel mode, this PSF structure due to the telescope is smeared by detector 
time constants and data averaging, resulting in a larger PSF. 
The PSFs delivered by the PACS instrument control centre (ICC) have been observed using the 
Vesta asteroid and provide information about the central part of the PSFs (until a radius of $\sim 60''$) at different scan speeds.
Mars observations were used for the characterisation of the encircled energy fraction (EEF) in the wings of PSFs but were not
combined with Vesta observations in the PSFs delivered.
%and provide information about different scan speeds in the central part of the PSFs , but not the wings.  
The goal of this Research Note is to combine Vesta and Mars dedicated observations to 
provide new estimates of the PACS PSFs with an unprecedented dynamic range ($\sim 10^{6}$) to permit the 
proper characterisation of the central part and the wings. 

This research note is organised as follows:  
in Section~\ref{sect:obs} we present the PACS observations,
give details on the data processing followed, and describe the method used to produce the final PSFs.
In Section~\ref{sect:analysis} we analyse the PSFs and compare them to extragalactic images
and, finally, in Section~\ref{sect:conclusions} we draw our conclusions.

\section{Vesta and Mars PACS data processing}
\label{sect:obs}

{\it Herschel} PACS dedicated PSF observations 
are scan maps centred on various objects taken at 70 (blue band), 100 (green band), and 160 (red band) $\mu$m.
The core of the PSF is best characterised observing faint objects (e.g. the asteroid Vesta),
while the wings of the PSF can only be seen in observations of bright objects (e.g. Mars).
Using a combination of images of bright and faint objects, it is therefore possible to obtain a good characterisation of the PACS PSFs.

\subsection{Observations and data reduction}

\begin{table}[t]
\caption{ObsIDs of the available observations of Vesta and Mars. 
%Each blue or green observation has a corresponding red channel counterpart.
}
%\vspace{.2cm}}
\label{table:obs_Vesta_Mars}      
\centering          
\begin{tabular}{l c c c c }
\hline\hline
Band & Scan speed & Vesta & Mars\\
\hline
\multirow{2}{*}{blue / red} & \multirow{2}{*}{20} & 1342195472 &- \\
& & 1342195473 & - \\
\hline
\multirow{4}{*}{blue / red} & \multirow{4}{*}{60} & - & 1342231157\\
& & 1342195470 & 1342231158\\
& &  1342195471& 1342231159\\
& & - & 1342231160\\
\hline
\multirow{2}{*}{green / red} & \multirow{2}{*}{20} & 1342195476 & - \\
& & 1342195477 & - \\
\hline
\multirow{4}{*}{green / red} & \multirow{4}{*}{60} & - & 1342231161\\
& &  1342195474& 1342231162\\
& & 1342195475 & 1342231163\\
& & - & 1342231164\\
\hline
\hline
\vspace{-1.cm}
\end{tabular}
\end{table}

We consider dual-band observations of Vesta and Mars (see obsIDs in Table~\ref{table:obs_Vesta_Mars}) taken at the three different 
PACS wavelength bands and with a scan speed of $20''/s$ and $60''/s$. 
The {\it Herschel} Interactive Processing Environment (HIPE; v.12.1.0; \citealt{2010ASPC..434..139O}) 
was first used to bring the raw Level-0 data to Level-1 using the PACS calibration
tree PACS\_CAL\_65\_0 and the pipeline scripts for Solar System Objects (SSO). 
Maps were then produced using Scanamorphos (v.24.0, \citealt{2013PASP..125.1126R}) with pixel sizes of $1''$
and rotated of the roll angle $\theta_{\rm RA} = 292.44^{\circ}$ and 108.6$^{\circ}$ 
clockwise (for Vesta and Mars observations, respectively) 
so as to have the Z-axis of the spacecraft point up.

Unfortunately, there are no sufficient dedicated observations in parallel mode that are useful for a correct 
characterisation of the PSF.
We bypassed this problem partly simulating parallel-mode observations at 20$''/s$ and 60$''/s$ from observations 
in standard mode.
This is performed by modifying the HIPE pipeline to average fluxes and coordinates of two consecutive frames of our data,
therefore mimicking the reduced sampling frequency of the detectors (for the blue and green bands only) in parallel mode (\citealt{Lutz}; private communication).

The observations considered have an array-to-map angle (ama) of $\pm 42.4^{\circ}$, i.e. scan directions form an angle of $\pm 42.4^{\circ}$
with respect to the Z-axis of the spacecraft.
This angle corresponds to the SPIRE ``magic'' angle\footnote{Adopting an ama of $\pm 42.4^{\circ}$ provides a 
good coverage for fully sampled maps in the three SPIRE bands.} (\citealt{SPIRE}) and it is always used for observations in parallel mode.
Observations in standard mode can be performed using an ama = $\pm42.4^{\circ}$ or $\pm 63^{\circ}$. 

The scanning direction does not affect the shape of the PSF for low scanning speeds (10 and $20''/s$).
On the contrary, Vesta images obtained at fast scanning speed ($60''/s$) present a clear elongation along the scanning direction.
The central region of Mars images is saturated and no elongation is observed; Mars observations at lower scan speeds are therefore
not required and images taken at 60$''/s$ are used for the characterisation of the faint wings at both slow and fast scan speeds.
In the {\it Herschel} Science Archive there is no PACS observation (apart from dedicated observations for PSFs analysis) at $60''/s$ in 
standard mode with an ama = $\pm63^{\circ}$, we therefore consider only PSFs with ama = $\pm 42.4^{\circ}$.
For the observations of Mars, the pixels centered on the source are heavily saturated, leading to significant 
trails. When using Scanamorphos \citep{PICC} these artifacts are greatly reduced, masking the affected regions.

\subsection{Merging observations}

%In order to characterise the PSF on a wide dynamic range we merge Vesta and Mars observations.
In order to produce PSFs for images obtained at low scan speeds (10 and $20''/s$),
we merge Vesta and Mars observations at $20''/s$ and $60''/s$, respectively, 
while we use Vesta and Mars observations at $60''/s$ for PSFs for high scan speeds ($60''/s$).
The same is done for the parallel mode, using the corresponding partly simulated data.

\begin{figure}[t]
\begin{center}
\includegraphics[trim=10cm 2cm 10cm -4mm , clip=1,width=0.43\textwidth]{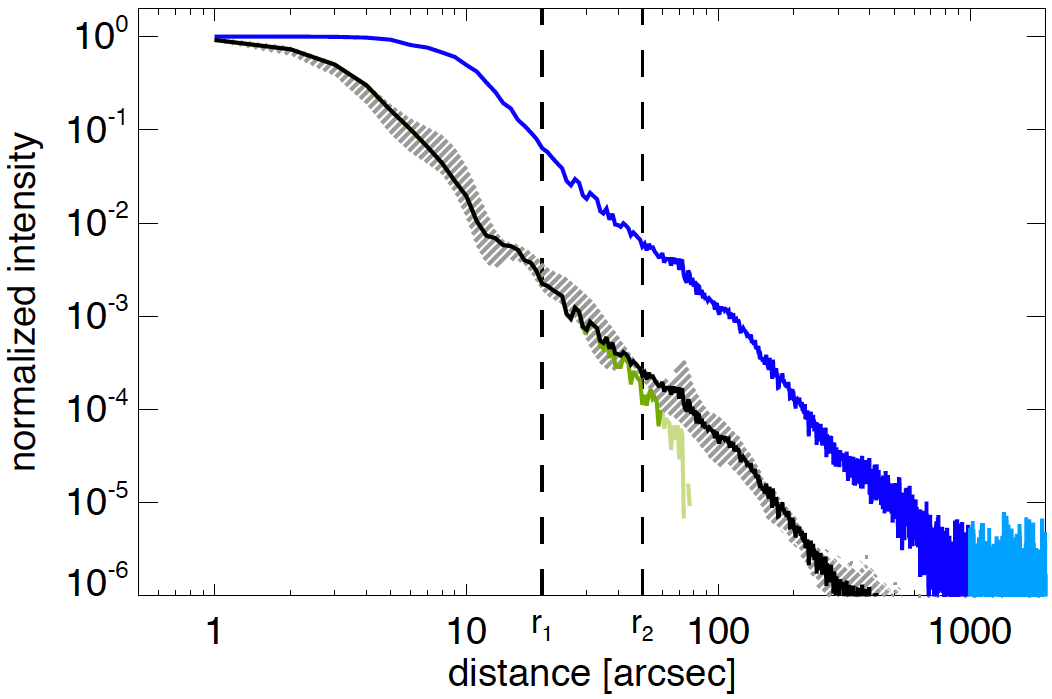}
\caption{Average radial profile of Vesta (green) and Mars (blue) observations and of the estimated 
PSF (black) for 70 $\mu$m images at a scan speed of $20''/s$. The light green and light blue lines are Vesta and Mars profiles 
for $r>c_2$. The shaded region indicates the range of
profiles along different directions of the estimated PSF. Vertical dashed black lines indicate $r_1$ and $r_2$.\vspace{-.2cm}}
\label{fig:prof}
\end{center}
\end{figure}

First of all, we notice that Mars images have a good signal-to-noise ratios (S/N) up to $\sim 1000''$ from the centre, while Vesta
images are noisy for radii that are larger than $60''$. We measure the background in Mars images and we remove it,
while the background estimation for Vesta is computed with a more sophisticated technique (see Sect.~\ref{sect:EEF}).
Images are then normalised so that the total integrated flux equals unity and the radial profile (averaged over the $2\pi$ angle) 
for both Vesta and Mars is computed (see Fig.~\ref{fig:prof} for an example at 70 $\mu$m with a scan speed of 20$''/s$). 
The central region of the Mars image is saturated
and the ratio between the two profiles is constant
over a given radial region (i.e. between $r_1 = 20''$ and $r_2 = 50''$ for 
observations in Fig.~\ref{fig:prof}).
%two profiles have approximately the same slope over a given radial region (i.e. between $r_1 = 20''$ and $r_2 = 50''$ for 
%observations in Fig.~\ref{fig:prof}). 
%We compute the average ratio between Vesta and Mars observations for $r_1 < r < r_2$.
To avoid introducing any artifacts in the PSFs, 
we then rescale the Mars images to those of Vesta and compute the PSF as
\begin{numcases}{PSF(r,\theta)=}
 V(r,\theta) & if $r \leq r_1$ \nonumber  \\
 V(r,\theta)[1-f(r)]+M(r,\theta)f(r) & if $r_1 < r < r_2$ \nonumber  \\
 M(r,\theta) & if  $r \geq r_2$,
\end{numcases}
where $V(r,\theta)$ and $M(r,\theta)$ indicate the Vesta and the rescaled Mars images, respectively, and
\begin{equation}
f(r) = 6 \left(\frac{r - r_1}{r_2 - r_1}\right)^5 - 15\left(\frac{r-r_1}{r_2-r_1}\right)^4+10\left(\frac{r-r_1}{r_2-r_1}\right)^3,
\end{equation} 
is a smooth Heaviside step function with null first and second derivatives at the extremes $r_1$ and $r_2$.

The average radial profile of the estimated PSF tightly follows that of Vesta at short radii, while it tends to the profile of Mars
further from the centre (see Fig.~\ref{fig:prof}).
During the operation of merging, the information on the asymmetry of the PSF is not lost and 
radial profiles measured along different directions show rather strong variability with respect to the average radial profile (shaded region in Fig.~\ref{fig:prof}).

\begin{figure*}[t]
\centering
\floatbox[{\capbeside\thisfloatsetup{capbesideposition={right,bottom},capbesidewidth=6cm}}]{figure}[\FBwidth]
{\begin{tikzpicture}[
squarednode/.style={rectangle,  fill=red!5, very thick, minimum height=1cm, minimum width=1cm,fill opacity=0.},
]
%draw=red!60,
 \node[anchor=south west,inner sep=0] at (0,0) {\includegraphics[trim=1.5cm 1cm .8cm 0cm,clip=1,width=0.65\textwidth]{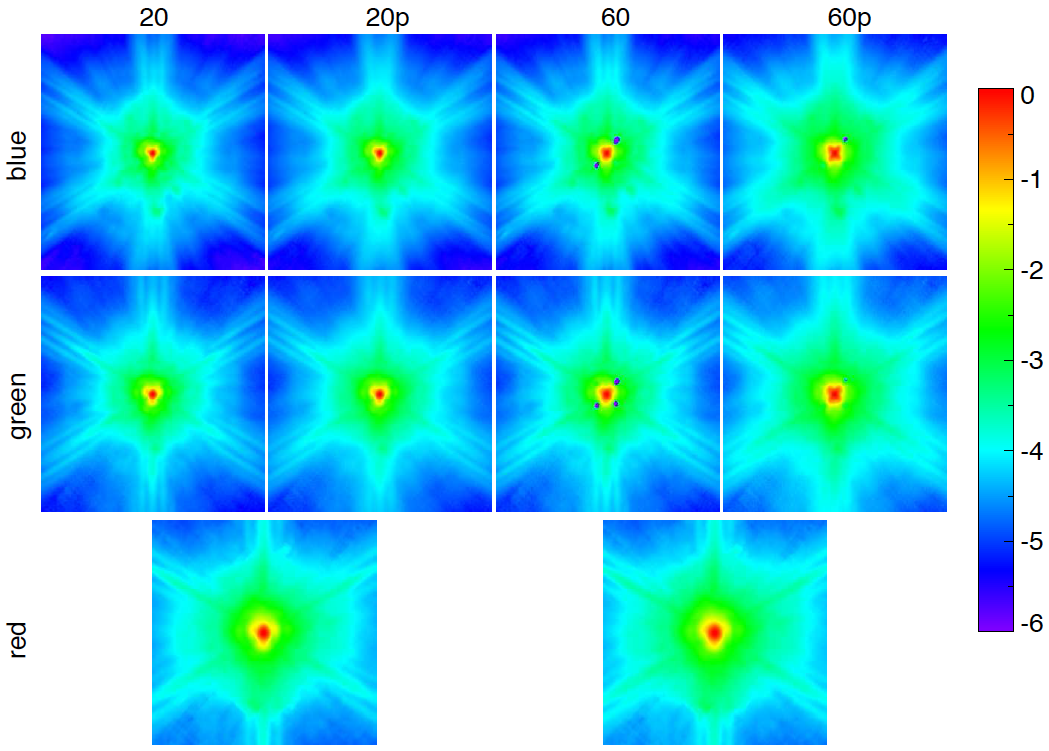}};
%  \node[squarednode] at (1.,1.35) {\href{http://idoc-herschel.ias.u-psud.fr/sitools/datastorage/user/storageRelease/R0_PSF/PACS/PHOTO/V20/Combined_PACS160_V20.fits}{\includegraphics[width=2.5cm,height=2.7cm]{images/PSF_PACS_mosaic_new.png}}};
%  \node[squarednode] at (3.6,1.35) {\href{http://idoc-herschel.ias.u-psud.fr/sitools/datastorage/user/storageRelease/R0_PSF/PACS/PHOTO/V20p/Combined_PACS160_V20p.fits}{\includegraphics[width=2.5cm,height=2.7cm]{images/PSF_PACS_mosaic_new.png}}};
%  \node[squarednode] at (6.25,1.35) {\href{http://idoc-herschel.ias.u-psud.fr/sitools/datastorage/user/storageRelease/R0_PSF/PACS/PHOTO/V60/Combined_PACS160_V60.fits}{\includegraphics[width=2.5cm,height=2.7cm]{images/PSF_PACS_mosaic_new.png}}};
%  \node[squarednode] at (8.85,1.35) {\href{http://idoc-herschel.ias.u-psud.fr/sitools/datastorage/user/storageRelease/R0_PSF/PACS/PHOTO/V20/Combined_PACS160_V60p.fits}{\includegraphics[width=2.5cm,height=2.7cm]{images/PSF_PACS_mosaic_new.png}}};
%  \node[squarednode] at (1,4.15) {\href{http://idoc-herschel.ias.u-psud.fr/sitools/datastorage/user/storageRelease/R0_PSF/PACS/PHOTO/V20/Combined_PACS100_V20.fits}{\includegraphics[width=2.5cm,height=2.7cm]{images/PSF_PACS_mosaic_new.png}}};
%  \node[squarednode] at (3.6,4.15) {\href{http://idoc-herschel.ias.u-psud.fr/sitools/datastorage/user/storageRelease/R0_PSF/PACS/PHOTO/V20p/Combined_PACS100_V20p.fits}{\includegraphics[width=2.5cm,height=2.7cm]{images/PSF_PACS_mosaic_new.png}}};
%  \node[squarednode] at (6.25,4.15) {\href{http://idoc-herschel.ias.u-psud.fr/sitools/datastorage/user/storageRelease/R0_PSF/PACS/PHOTO/V60/Combined_PACS100_V60.fits}{\includegraphics[width=2.5cm,height=2.7cm]{images/PSF_PACS_mosaic_new.png}}};
%  \node[squarednode] at (8.85,4.15) {\href{http://idoc-herschel.ias.u-psud.fr/sitools/datastorage/user/storageRelease/R0_PSF/PACS/PHOTO/V60p/Combined_PACS100_V60p.fits}{\includegraphics[width=2.5cm,height=2.7cm]{images/PSF_PACS_mosaic_new.png}}};
%  \node[squarednode] at (1,6.95) {\href{http://idoc-herschel.ias.u-psud.fr/sitools/datastorage/user/storageRelease/R0_PSF/PACS/PHOTO/V20/Combined_PACS70_V20.fits}{\includegraphics[width=2.5cm,height=2.7cm]{images/PSF_PACS_mosaic_new.png}}};
%  \node[squarednode] at (3.6,6.95) {\href{http://idoc-herschel.ias.u-psud.fr/sitools/datastorage/user/storageRelease/R0_PSF/PACS/PHOTO/V20p/Combined_PACS70_V20p.fits}{\includegraphics[width=2.5cm,height=2.7cm]{images/PSF_PACS_mosaic_new.png}}};
%  \node[squarednode] at (6.25,6.95) {\href{http://idoc-herschel.ias.u-psud.fr/sitools/datastorage/user/storageRelease/R0_PSF/PACS/PHOTO/V60/Combined_PACS70_V60.fits}{\includegraphics[width=2.5cm,height=2.7cm]{images/PSF_PACS_mosaic_new.png}}};
%  \node[squarednode] at (8.85,6.95) {\href{http://idoc-herschel.ias.u-psud.fr/sitools/datastorage/user/storageRelease/R0_PSF/PACS/PHOTO/V60p/Combined_PACS70_V60p.fits}{\includegraphics[width=2.5cm,height=2.7cm]{images/PSF_PACS_mosaic_new.png}}};
\end{tikzpicture}}
{\caption{Our estimates of PACS PSFs (in log scale) as a function of the filter and scan speed. All images are $300'' \times 300''$.
Red band PSFs in standard and parallel modes are exactly the same. The spacecraft Y- and Z-axis are to the left and to the top, respectively.
%A clickable version of the image is available online. 
}
\label{fig:mosaic}}
\end{figure*}

We also tested for the impact of the finite size of Mars on the
PSF determination. At the time of observations, the apparent diameter
of the planet as seen from the L2 point\footnote{Ephemerides can be accessed at: \url{http://ssd.jpl.nasa.gov/?horizons}} was $d_{\rm M}\sim 5\farcs55$, 
which is comparable to the FWHM of the PACS PSFs and therefore cannot be considered point-like. However, the core of PSFs is obtained from
Vesta measurements and Mars observations are only used for the characterisation of the faint structures at $r>r_1$. 
When a circle of diameter $d_{\rm M}\sim 5\farcs55$ is convolved with our composite PSFs,
the profile of the resulting images for $r>r_1$ remains unchanged
with respect to the PSFs. This demonstrates that the finite size
of Mars does not affect the shape of the derived PSFs.

\section{PSF analysis}
\label{sect:analysis}
 
\subsection{Encircled energy}
\label{sect:EEF}

The EEF is computed using the same notation and following the method by the ICC (\citealt{PICC}) as follows:
\begin{equation}
EEF_{\rm obs}(r) = \frac{\int_0^r \int_0^{2\pi} [PSF(r,\theta)-c_1]\,rdr\,d\theta - c_3}{\int_0^{c_2} \int_0^{2\pi} [PSF(r,\theta)-c_1]\,rdr\,d\theta - c_3},
\end{equation}
where $c_1$ represents the background value to be removed from the observed image, $c_2$ is the maximum radius out to which we compute
the EEF, and $c_3$ is the flux missing in the PSF core due to saturation.

\begin{figure}[b]
\begin{center}
\includegraphics[trim=10cm 2cm 10cm -4mm , clip=1,width=0.43\textwidth]{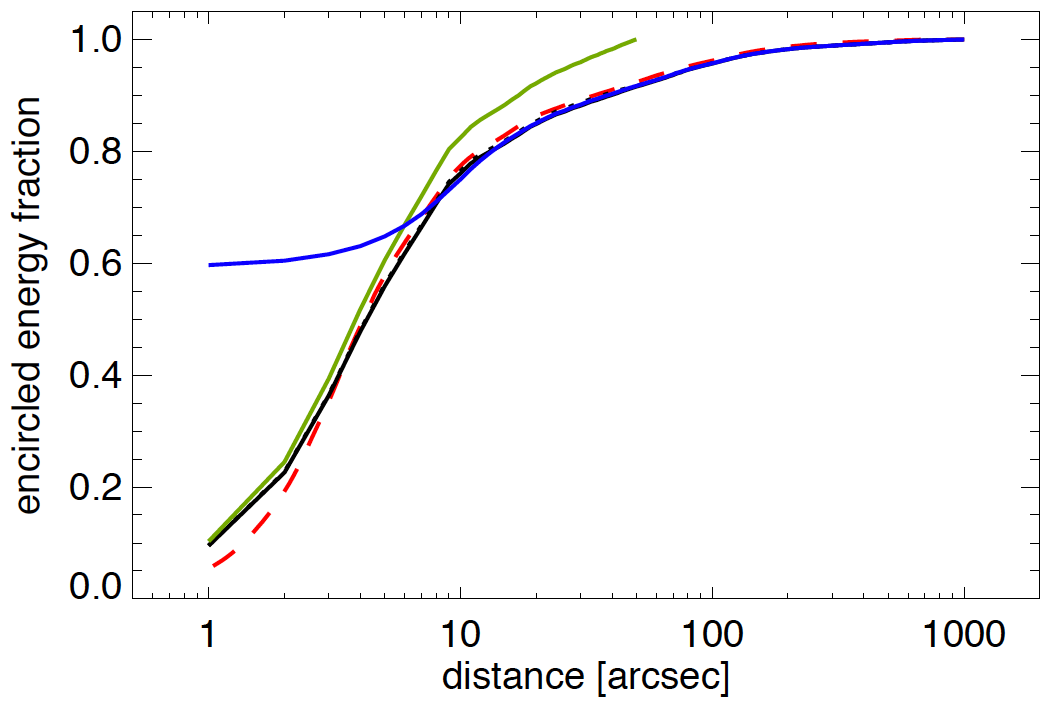}
\caption{Encircled energy fraction (blue band) for Vesta (green line), Mars (blue line), for the estimated PSF (solid black line), 
and as obtained by the ICC (red dashed line).
The dotted black line indicates the corrected Vesta EEF (see text for details).}
\label{fig:EEF}
\end{center}
\end{figure}

We compute the EEF for the Vesta and Mars images (see Fig.~\ref{fig:EEF}) assuming $c_1 = 0$, $c_2 = 60''$ and $1000''$ for Vesta and Mars, respectively.
We deduced the value of $c_3$ for Mars images  by comparing  the integrated observed flux to the nominal flux given by the ICC 
(i.e. $S[70] = 44670$ Jy, $S[100] = 24740$ Jy and $S[160] = 11160$ Jy).
The observed fluxes are $\sim 37.5\%, 53.6\%$, and $86.8\%$ of the nominal values at 70, 100, and 160 $\mu$m, respectively.

\begin{table}[b]
\caption{FWHM of the estimated PSFs in arcsec along the Y and Z directions. 
Scan speeds are indicated in arcsec/s, 20p and 60p are in parallel mode.\vspace{.0cm}}
\label{table:FWHM}      
\centering          
\begin{tabular}{@{\extracolsep{1mm}} l c c c c c }
\hline\hline
\multirow{2}{*}{Band} &\multirow{2}{*}{Dir.}& \multicolumn{4}{c}{Scan speed}\\
\cline{3-6}
& & 20& 20p & 60 & 60p\\
\hline
\noalign{\vskip 3pt}  
\multirow{2}{*}{blue}  &\multicolumn{1}{|c}{Y}  & 5.77   &  6.28    &  7.75  &   9.71  \\
                                   &\multicolumn{1}{|c}{Z} & 6.35   & 6.94    & 8.72    & 10.77 \vspace{2pt}\\
\multirow{2}{*}{green}&\multicolumn{1}{|c}{Y} & 6.90   & 7.31     & 8.68    & 10.67  \\
                                  &\multicolumn{1}{|c}{Z} & 7.26   & 7.75     & 9.51    &  11.72\vspace{2pt}\\
\multirow{2}{*}{red}     &\multicolumn{1}{|c}{Y}& \multicolumn{2}{c}{10.59}    &  \multicolumn{2}{c}{11.80}   \\
                                  &\multicolumn{1}{|c}{Z}& \multicolumn{2}{c}{12.29}& \multicolumn{2}{c}{13.70}\vspace{2pt}\\
\hline
\hline
\end{tabular}
\end{table}
However, the flux of Vesta for radii larger than $c_2$ is non-negligible and must be taken into account for 
the calculation of the EEF.
Furthermore, since the S/N in Vesta maps is very low at $r > c_2$, the value of $c_1$ cannot be estimated
directly from observations.
From a visual comparison of the slope of the Vesta and Mars EEF, we find $c_1 \simeq 10^{-4}$ (normalised to the peak of the Vesta image) 
for all pairs of observations.
The flux outside radius $c_2$ in Vesta observations is 8.3\%, 9.8\%, and 12.3\% for blue, green, and red bands.
We then correct the Vesta EEF for the flux lost due to the non-zero $c_1$ and the flux outside radius $c_2$ (dotted line in Fig.~\ref{fig:EEF}).
The corrected EEF curve tightly follows that of Vesta at short radii and tends to that of Mars for $r \gtrsim 20''$.
We finally compute the EEF for our estimate of the PACS PSF (solid line in Fig.~\ref{fig:EEF}) and note that it agrees very well
with the corrected Vesta EEF, therefore supporting the methodology used.
The EEF curve that we obtain is comparable to that presented by the PACS ICC (red dashed line in Fig.~\ref{fig:EEF}, \citealt{PICC}),
with a little difference for $r \lesssim 2''$ due to the adopted centering technique.
\begin{figure*}[t]
\begin{center}
\floatbox[{\capbeside\thisfloatsetup{capbesideposition={right,bottom},capbesidewidth=6cm}}]{figure}[\FBwidth]
{\includegraphics[clip=1,trim=10cm 6cm 2cm 0cm,width=0.58\textwidth]{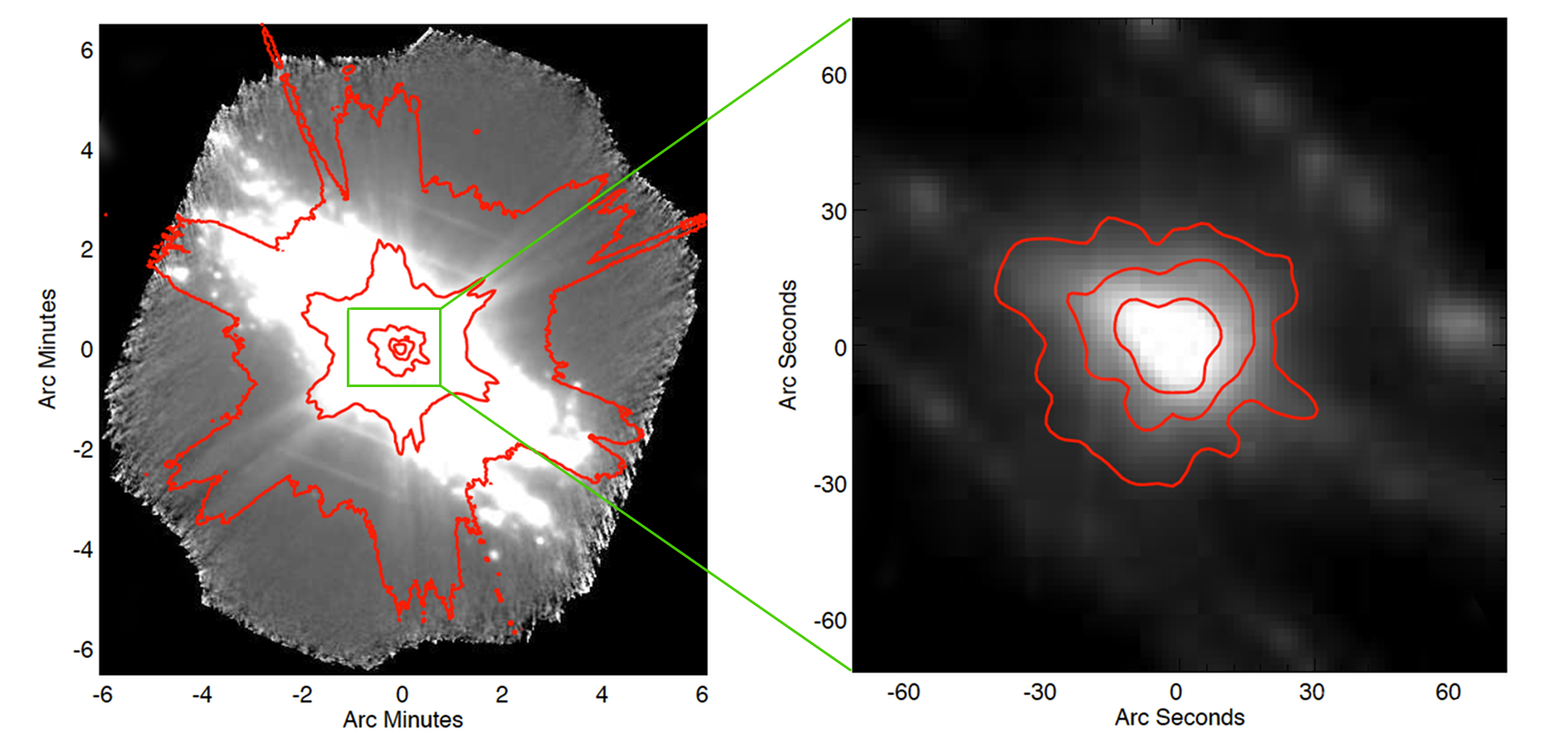}\vspace{-.5cm}}
{\caption{PACS 70 $\mu$m observations of NGC 253 with overplotted contours of our PSF, rotated of an angle $\theta_{\rm RA} \sim 114^{\circ}$ clockwise (see text).
Right panel, zoom to the central region.\vspace{-.5cm}}
\label{fig:NGC253}}
\end{center}
\end{figure*}

%I got it, the value of c1 is estimated from the slope of the EEF curves.
% the value of c1 does change the slope, for example the slope for Vesta EEF (c1 = 0) is different from the corrected one (c1 = 1.d-4).

\subsection{PSF profile}
The computed PACS PSFs are presented in Fig.~\ref{fig:mosaic}.
% and are downloadable from \url{http://idoc-herschel.ias.u-psud.fr/sitools/client-user/Herschel/project-index.html}.
They are not axisymmetric and are characterised by a large width variability depending on the direction.
Using a two-dimensional (2D) Gaussian fit of the PSFs, we measure the FWHM along the Y and Z directions.
The resulting values are reported in Table~\ref{table:FWHM} and are comparable to those obtained by the PACS ICC (\citealt{PICC}).
As expected, the FWHM increases in both Y and Z directions from blue to red filters.
PSF features are observed along the Z direction (see Fig.\ref{fig:mosaic}) and the width is therefore systematically larger (up to 20\%) along this axis
with respect to the Y direction.
The scan speed contributes to the enlargement of the PSF and a clear elongation is visible towards the scanning direction ($\pm 42.4^{\circ}$ with respect to 
the Z axis).

\subsection{An example}

Astronomical objects presenting a strong contrast between different regions can present evident PSF features.
NGC 253 is a intermediate spiral galaxy currently undergoing an intense star formation.
This galaxy has been observed by PACS at 70 and 160 $\mu$m at a scan speed of $20''/s$ and with the $+Z$ direction of the telescope rotated 
by an angle $\theta_{\rm RA} \sim 114^{\circ}$ west of north (clockwise).
At these wavelengths, the central region of the galaxy is very bright compared to the rest of the galaxy, which is then contaminated by
the lobes and faint structures of the PSFs.

Fig.~\ref{fig:NGC253} illustrates the PACS 70 $\mu$m image of this galaxy with  the contours of our estimated PSF overplotted.
On the right panel, we show a zoom to the central region and overplot contours of the PSF core.
Both the faint structures and core of the PSF dominate over the extended emission of the galaxy and match very well with the 
PSF\footnote{The faint stripes visible in the left panel are due to crosstalk (\citealt{okumura}) and are not directly related to the shape of the PSF.}.
This observation represents an example of a case where a good characterisation of the PSF of the instrument is needed to 
correctly interpret astrophysical data.

Similarly, our computed PACS PSFs were used in a recent study of the scale height of the dust distribution 
in a nearby edge-on galaxy, NGC 891 (\citealt{2016A&A...586A...8B}).
The larger width of the PSFs compared to that of modelled PSFs for the `as built' telescope (\citealt{PICC29})
and their radial asymmetry lead to a narrower dust scale height
by up to a factor of $\sim 60\%$.

\section{Conclusions}
\label{sect:conclusions}

Using dedicated observations of Vesta and Mars, we provide new estimates of the PACS PSFs for scan speeds of 20$''/s$ and 60$''/s$ both in standard
and parallel mode.
The obtained PSFs have a wide dynamic range ($\sim 10^6$) enabling a proper characterisation of both the core and faint structures
of the PSFs.

As an example we consider NGC 253, a galaxy with a strong contrast between the central and peripheral regions.
From a comparison between our estimated PSFs and PACS observations of NGC 253, we obtain an excellent matching, therefore
supporting the reliability of the method used.

%The presented PSFs are publicly available and downloadable from \url{http://idoc-herschel.ias.u-psud.fr/sitools/client-user/Herschel/project-index.html}.

\begin{acknowledgements}
We would like to acknowledge Prof. D. Lutz for a useful discussion and for giving us part of the code needed to simulate parallel-mode observations.
We thank the PACS ICC for their insightful comments on the paper.
We acknowledge K. Dassas for making the PSFs available online on the IDOC website.
%at \url{http://idoc-herschel.ias.u-psud.fr/sitools/client-user/Herschel/project-index.html}.
Part of this work has received funding from the European
UnionÕs Seventh Framework Programme (FP7/2007-2013) for the DustPedia
project (grant agreement n$^{\circ}$ FP7-SPACE-606847).
\end{acknowledgements}

\bibliographystyle{template/aa}
\bibliography{bib_PSFs}{}

\end{document}